\documentclass{jps-cp}
\usepackage{graphicx}  
\usepackage{dcolumn}   
\usepackage{bm,relsize}        
\usepackage{amssymb, amsmath}
\usepackage{textcomp}
\usepackage{wasysym}
\usepackage{slashed}
\usepackage{multirow}

\usepackage{array}
\newcolumntype{H}{>{\setbox0=\hbox\bgroup}c<{\egroup}@{}}

\usepackage{lipsum, color}
\usepackage[usenames,dvipsnames,svgnames]{xcolor}
\usepackage{braket}
\usepackage{pdflscape}

\usepackage{makecell}

\newcommand{\CO}{\mathcal{O}}

\def\beq{\begin{equation}}
\def\eeq{\end{equation}}
 \def\be{\begin{equation}} \def\ee{\end{equation}}
\def\bea{\begin{eqnarray}} \def\eea{\end{eqnarray}}

\usepackage{multirow}
\usepackage[normalem]{ulem}

\allowdisplaybreaks

\title{The Effective Theory of $\mu \to e$ Conversion}

\author{W. C. Haxton$^{1,2,3}$ and Evan Rule$^{1,4}$}

\inst{$^{1}$Department of Physics, University of California, Berkeley, CA 94720, USA \\
$^{2}$Institute for Nuclear Theory, University of Washington, Seattle, WA 98195 \\
$^{3}$Lawrence Berkeley National Laboratory, Berkeley, CA 94720, USA \\
$^{4}$Theoretical Division, Los Alamos National Laboratory, Los Alamos, NM 87545, USA}

\email{haxton@berkeley.edu, erule@lanl.gov} 

\recdate{January 11, 2026}

\abst{We summarize recent work to develop an effective theory of $\mu \rightarrow e$ conversion, based on a complete set of low-energy effective operators that are developed from a systematic expansion in velocities and momenta. The expansion effectively factors rates into sums of particle physics and nuclear physics terms, where the former are expressed as bilinears in the LECs (the low-energy constants of the  effective theory) and the latter are the associated nuclear responses.  One can view the nuclear responses as ``dials" that can be adjusted --- for example, by selection of targets with specific properties --- in order to isolate the former.  We show that an important dial, in the case of Mu2e and COMET, will be inelastic transitions to certain low-energy nuclear states that are resolvable in $^{27}$Al.  If these transitions are exploited, the experiments have the potential not only to discover charged lepton flavor violation (CLFV), but to determine the operators responsible for the CLFV.  We also discuss how such low-energy results can be ``ported" to higher energies through a tower of matched EFTs, so they can be combined with other experimental limits to further constrain CLFV.}

\kword{lepton flavor, muon-to-electron conversion, charged lepton flavor violation}

\begin{document}
\maketitle

\section{Introduction}
The discovery of neutrino oscillations \cite{Super-Kamiokande,SNO} established that lepton flavor is not only violated in interactions among neutral leptons, but also among the charged leptons, due to the charged-lepton flavor violation (CLFV) induced by neutrino loop corrections.  However, this source of CLFV is suppressed by the tiny scale of neutrino masses, yielding rates too small to be observed experimentally. 
Consequently, any detection of CLFV would be evidence of new physics. There are many plausible modifications of the standard model (SM) --- including supersymmetry, leptoquarks, heavy neutrinos, or a more complicated Higgs sector --- that could induce observable levels of CLFV.  Current upper bounds on CLFV branching ratios thus provide stringent constraints on such UV BSM extensions \cite{BWLee,Calibbi:2017uvl,Calibbi}.

Among the most sensitive CLFV measurements are those performed using stopped muons. The current best limits on the CLFV branching ratios $B(\mu^+\rightarrow e^+\gamma)<1.5\times 10^{-13}$ and $B(\mu^+\rightarrow e^+e^-e^+)<1.0\times 10^{-12}$ were obtained at 90\% confidence level (C.L.) by MEG II \cite{MEGII:2025gzr} and SINDRUM \cite{SINDRUM:1987nra}, respectively. The former bound is expected to improve by another factor of two due to continued data collection at MEG II \cite{MEGII:2018kmf}, while the latter bound will be tightened by at least three orders of magnitude by the Mu3e experiment \cite{Mu3e:2020gyw}. 

However, as described in talks presented at this meeting, a major advance in sensitivity to CLFV is expected in the next few years due to new experiments on muon-to-electron conversion, $\mu^-+(A,Z)\rightarrow e^- + (A,Z)$.  This process can be observed by stopping muons in a thick target, where they capture into the first Bohr orbit of the target nucleus, then searching for electrons emitted with an energy of $m_\mu$ (neglecting small corrections due to atomic binding and nuclear recoil).  Mu2e and COMET, under construction at Fermilab and J-PARC respectively, are expected to improve limits on the $\mu \rightarrow e$ rate by approximately four orders of magnitude. Given the many CLFV mechanisms that have been suggested in the literature, several questions naturally arise: As these measurements are done at low energies, what information can they provide on the BSM source(s) of CLFV?  Is it just one rate that can be measured,
or can multiple constraints be obtained by changing target properties, etc?  In other words, in principle, how many $\mu \rightarrow e$ particle physics observables exist, and what can be done to separately determine each observable?  How does one ensure that the analysis framework one uses to interpret an experimental limit is complete, to avoid being misled by one's assumptions?  Finally, assuming one has properly analyzed an experiment, what does the extracted constraint (or constraints) imply for the BSM source of the CLFV, which presumably operates at a much higher energy scale?  How does one relate the measurement to 
possible BSM theories of CLFV, or alternatively, to experimental CLFV limits from other processes, including those obtained at collider energies?  

There is a theory tool ideal for this situation --- effective field theory. An EFT formulation appropriate for $\mu \rightarrow e$ conversion has been
developed by us over the past three years, in collaboration at different times with Ken McElvain, Tony Menzo, Michael Ramsey-Musolf, and Jure Zupan, who we gratefully acknowledge.  Because the work is published \cite{ER1,ER2,Haxton:2024ecp,tower,PhysRevC.111.025501}, both the original presentation at SSP2025 and this summary are designed as overviews, with the references available to readers interested in more detailed information.

\section{Muon-to-Electron Conversion}
The quantity extracted from experiment is the branching ratio for $\mu \rightarrow e$ conversion
\begin{equation}
\label{eq:B:mu2e}
B(\mu^-\rightarrow e^-)=\frac{\Gamma\left[\mu^-+(A,Z)\rightarrow e^-+(A,Z)\right]}{\Gamma\left[\mu^-+(A,Z)\rightarrow \nu_\mu +(A,Z-1)\right]},
\end{equation}
with respect to ordinary muon capture in a nucleus with $A$ nucleons and $Z$ protons.  SM muon capture rates are generally available for targets of interest, due to the long history of measurements \cite{PhysRevC.35.2212}. The most stringent current bounds on $\mu \rightarrow e$ branching ratios are $B(\mu^-+$Ti$\rightarrow e^-+$Ti$)<6.1 \times 10^{-13}$ \cite{Wintz:1998rp} and $B(\mu^-+$Au$\rightarrow e^-+$Au$)<7 \times 10^{-13}$ \cite{SINDRUMII:2006dvw}. Both of these 90\% C.L. limits were established by the SINDRUM II collaboration. By the end of this decade, the new Fermilab (Mu2e \cite{Mu2e:2014fns,Bernstein:2019fyh}) and J-PARC (COMET \cite{COMET:2018auw,COMET:2018wbw}) experiments are expected to reach branching ratio sensitivities of about 10$^{-17}$, probing new physics scales in excess of $10^4$ TeV. Both experiments will measure electrons produced from the conversion of a muon bound in the $1s$ atomic orbital of $^{27}$Al. 

The properties of nuclear targets --- the nucleon number $A$ and proton number $Z$, the ground-state spin and isospin, and their responses to operators that involve orbital angular momentum $\vec{\ell}$, spin $\vec{s}$, or spin-orbit correlations $\vec{\ell} \cdot \vec{s}$ --- affect the relationship between $\mu \rightarrow e$ conversion bounds and underlying sources of CLFV. Comparable branching ratio limits obtained from different nuclear targets will differ in their sensitivity to a given source of CLFV.  This is an attractive feature of $\mu \rightarrow e$ conversion that distinguishes this process from $\mu \rightarrow e \gamma$ or $\mu \rightarrow 3e$: by using several nuclear targets, one can place multiple constraints on the CLFV mechanism.  Were $\mu \rightarrow e$ conversion to be discovered in a given target, one could conduct additional studies with suitably selected targets to further characterize the CLFV source.

On the other hand, the interpretation of experimental limits on $\mu\to e$ conversion is, compared to $\mu\rightarrow e\gamma$ and $\mu\rightarrow 3e$, significantly more complicated due to the use of nuclear targets. Although the literature on $\mu\rightarrow e$ conversion is extensive (see the summary in \cite{ER2}), most existing work has focused on the special case of coherent conversion, where the leading response is governed by a scalar, isoscalar operator that sums coherently over all nucleons in the target. With this operator choice, the nuclear physics simplifies dramatically, allowing one to compute the $\mu\rightarrow e$ conversion rate using experimentally determined proton and neutron densities (see, e.g., Ref. \cite{Kitano:2002mt}).

\begin{figure}[t]
    \centering
    \includegraphics[width=0.6\textwidth]{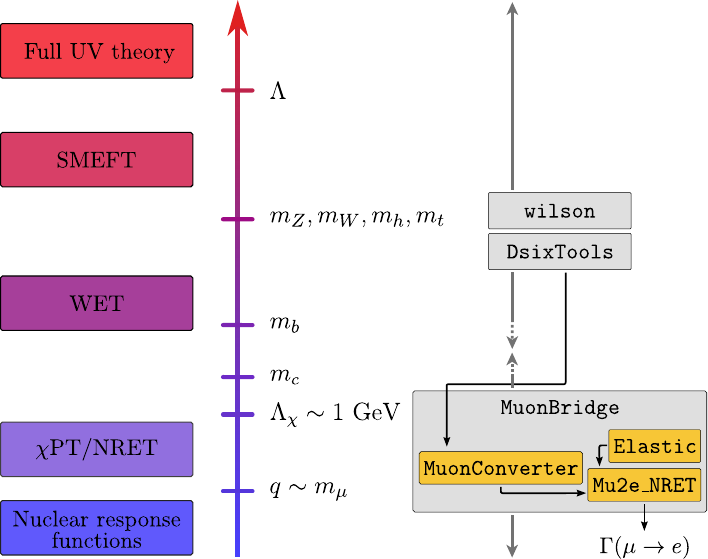}
    \caption{The tower of linked effective field theories that can be employed to connect low-energy $\mu \rightarrow e$ results to higher energy scales. The new \texttt{MuonBridge} software is described in \cite{tower}.}
    \label{fig:tower}
\end{figure}

While studies of the coherent operator have provided useful CLFV benchmarks~\cite{Cirigliano:2009bz,Cirigliano:2022ekw,Ardu:2024bua}, symmetry arguments show that six nuclear response functions arise in a general description of elastic $\mu \rightarrow e$ conversion \cite{ER1,ER2}. Given that we have no a priori information on the CLFV mechanism, experiments should be interpreted using a formalism that takes into account all possible low-energy CLFV effective interactions.
The situation is reminiscent of another elastic process, the elastic scattering of WIMP dark matter (DM) off a nucleus.  Over the last decade, an attractive formalism was developed for this process in which DM direct detection limits are analyzed using an effective theory formulated for the low-energy nuclear scale, where experiments are performed.  The starting point is a Galilean-invariant nonrelativistic effective theory (NRET)  \cite{Fan:2010gt,Fitzpatrick:2012ix,Anand:2013yka} consisting of all possible heavy DM-nucleon interactions that can be constructed from the available hermitian operators. This guarantees that all operators allowed by symmetry --- and thus the full set of nuclear response functions --- will be generated. The NRET is both complete, including all operators of a given order, and efficient, as the number of degrees of freedom
are minimized by working at the energy scale of the experiments. The NRET is then connected to UV theories of DM through a tower of effective field theories that are matched at their interfaces \cite{Bishara:2018vix,Baumgart:2022vwr}. The papers summarized here \cite{ER1,ER2,Haxton:2024ecp,tower,PhysRevC.111.025501} 
developed an analogous formalism for $\mu \rightarrow e$ conversion, including both elastic and inelastic interactions with the nucleus, completing the EFT tower depicted in Fig. \ref{fig:tower}.

\section{Nonrelativistic effective theory}
\label{sec:NRET}
The NRET is constructed at the nucleon level by combining the available Hermitian operators to form all allowed scalars. The available operators include the intrinsic velocities of the nucleon $\vec{v}_N$ and muon $\vec{v}_{\mu}$ as well as the recoil velocity of the nuclear target $\vec{v}_T$. ``Intrinsic'' means these velocities are Galilean invariant, describing the relationship of the muon to the nuclear center-of-mass, and of the nucleons relative to each other (i.e., the $A-1$ internucleon Jacobi velocities).  These velocities satisfy the hierarchy $\vec{v}_N > \vec{v}_\mu \gg \vec{v}_T$. Consequently the target velocity is ignored. The full NRET retains operators linear in either $\vec{v}_N$ or $\vec{v}_\mu$. However, $\vec{v}_\mu$, which is associated with the bound muon's lower component, does not play a role in the nuclear selection rules that govern the form of the nuclear rate: one finds that its net effect in the NRET is to modestly modify nuclear form factors, typically changing results by $\approx 5\%$.  Thus to simplify the presentation, we will neglect $\vec{v}_\mu$. This reduces the available Hermitian operators to four: $i \hat{q}$ where $\hat{q}$ is the direction of the outgoing electron, the nucleon velocity $\vec{v}_N$, and the respective lepton and nucleon spin operators, $\vec{\sigma}_L$ and $\vec{\sigma}_N$. 
From these operators one can construct 16 independent CLFV single-nucleon interactions
\begin{align}
\label{eq:ops}
\CO_1 &= 1_L ~1_N, & \CO^\prime_2 &= 1_L ~i \hat{q} \cdot \vec{v}_N,  \notag \\
 \CO_3 &= 1_L~  i \hat{q} \cdot  \left[ \vec{v}_N \times \vec{\sigma}_N \right], & 
  \CO_4 &= \vec{\sigma}_L \cdot \vec{\sigma}_N,  \notag \\
\CO_5 &=  \vec{\sigma}_L \cdot \left( i \hat{q} \times \vec{v}_N \right), &
 \CO_6&=  i \hat{q} \cdot \vec{\sigma}_L ~ i \hat {q} \cdot \vec{\sigma}_N,  \notag \\
\CO_7 &= 1_L~ \vec{v}_N \cdot \vec{\sigma}_N, &
\CO_8 &= \vec{\sigma}_L\cdot \vec{v}_N, \notag \\ 
\CO_9 &= \vec{\sigma}_L \cdot \left( i \hat{q} \times \vec{\sigma}_N \right), &
\CO_{10} &= 1_L~ i \hat{q} \cdot \vec{\sigma}_N, \notag \\
\CO_{11} &= i \hat{q} \cdot \vec{\sigma}_L ~ 1_N,  &
\CO_{12} &= \vec{\sigma}_L \cdot \left[ \vec{v}_N \times \vec{\sigma}_N \right], \notag \\ 
\CO^\prime_{13} &= \vec{\sigma}_L \cdot  \left( i \hat{q} \times \left[ \vec{v}_N \times \vec{\sigma}_N \right] \right),  &
\CO_{14} &= i  \hat{q} \cdot \vec{\sigma}_L ~ \vec{v}_N \cdot \vec{\sigma}_N,  \notag  \\
\CO_{15} &= i \hat{q} \cdot \vec{\sigma}_L~ i \hat{q} \cdot \left[ \vec{v}_N \times \vec{\sigma}_N \right], &
\CO^\prime_{16} &= i \hat{q} \cdot \vec{\sigma}_L ~i \hat{q} \cdot  \vec{v}_N. 
\end{align}
This operator basis matches closely the one previously derived for dark matter direct detection \cite{Fitzpatrick:2012ix}: we distinguish with a prime the operators for which there are significant differences.
The NRET operators $\CO_i$ are understood to act between Pauli spinors $\xi_s$ for the muon, electron, and nucleon. The leptonic current operators $1_L$ or $\vec{\sigma}_L$ couple the $1s$ muon wave function to the various distorted partial waves comprising the outgoing electron's wave function.

The corresponding effective interaction can be expressed as 
\begin{equation}
    \mathcal{L}_\mathrm{eff}^{\rm NRET}= \sum_{\tau=0,1} \sum_{i=1}^{16} c_i^{\tau} \mathcal{O}_i t^{\tau}+\cdots,
    \label{eq:L_NRETv2}
\end{equation}
where ellipses denote $\vec{v}_\mu$-dependent terms, interactions quadratic in $\vec{v}_N$,  and other higher-order corrections (including, when embedded in the nucleus, two-nucleon terms). The numerical coefficients $c_i^\tau$, the low-energy constants (LECs) of the NRET, are the Wilson coefficients for the NRET and carry dimensions of $1/(\mathrm{mass})^{2}$. In principle,the LECs are functions of $\vec q^{\;2}$, though for $\mu \to e$ conversion this is a fixed quantity $\sim m_\mu$, apart from small nuclear-target-dependent corrections. Consequently, the NRET LECs are genuine constants capable of absorbing the momentum dependence that arises, for example, from the exchange of axion-like or other light mediators with $m^2\lesssim \vec{q}^{\;2}$.  An isospin index $\tau$ is included in Eq. \eqref{eq:L_NRETv2} to allow for any combination of isoscalar and isovector couplings to the nucleon, with $t^0=1$ and $t^1=\tau_3$ the isospin matrices.  Alternatively, one can introduce independent couplings to the proton and neutron, in which case $c_i^0=(c_i^p+c_i^n)/2$ and $c_i^1=(c_i^p-c_i^n)/2$.

The embedding of Eq. (\ref{eq:L_NRETv2}) in a nucleus is not entirely trivial.  First, $\vec{v}_N$ must be treated as an operator, acting on the relative coordinates of bound nucleons.  Second, the contact form of the NRET interaction leads to a transition amplitude formed from the convolution of leptonic and nuclear densities.  If one neglects the muon's lower component, replaces the upper component (which varies slowly over the nucleus) by its average value, and neglects Coulomb effects on the outgoing electron, the nuclear operators would be obtained by
\begin{equation}
        \mathcal{O}_i t^{\tau} \rightarrow \sum_{j=1}^A e^{i \vec{q} \cdot \vec{r}_j} \,\mathcal{O}_i (j)\, t^{\tau}(j)
    \label{eq:L_NRETv3}
\end{equation}
The nucleus cannot be treated as a point target, as $q$ is comparable to the inverse nuclear size; in a harmonic-oscillator single-particle nuclear basis,
the nuclear size effects are encoded in the dimensionless parameter $y \equiv (q b/2)^2 \approx \frac{1}{2}$, where $b$ is the oscillator
parameter.  As $y$ is not small, a partial wave expansion must be done to properly account for the
angular momentum transfer from leptons to the nucleus.  As the Coulomb effects can be
substantial, the expansion should be done with distorted electron waves. 

This task is sufficiently challenging that the partial-wave physics has been ignored in most past work, even though $y$ is by far the largest of the ``small parameters" that should guide expansions. However,
one can incorporate the effects of the distorted Dirac electron waves while retaining the 
simplicity of plane waves through the replacement \cite{ER1,ER2}
\begin{equation}
    U(q,s) e^{i \vec{q} \cdot \vec{r}} \rightarrow \frac{q_\mathrm{eff}}{q} \sqrt{\frac{E_e}{2m_e}} \, \left( \begin{array}{c} \xi_s \\ \vec{\sigma}_L \cdot \hat{q} \xi_s  \end{array}\right) e^{i \vec{q}_\mathrm{eff} \cdot \vec{r}}.
\end{equation}
where the effective momentum $\vec{q}_\mathrm{eff}$ is the electron's local momentum in the Coulomb potential of the nucleus. This replacement accurately accounts for Coulomb wavelength and amplitude shifts, yet allows one to perform a standard Fourier-Bessel partial-wave expansion of the nuclear rate.  One finds \cite{ER1,ER2}
\beq
\begin{split}\label{eq:Gamma:mutoe}
\Gamma(\mu\to e)=\frac{1}{2 \pi}\frac{q_{\rm eff}^2}{1+q/M_T}&\big|\phi_{1s}^{Z_\mathrm{eff}}(\vec{0})\big|^2\\
\times\sum_{\tau,\tau'}\Bigg\{&\Big[R_{M}^{\tau\tau'} W_{M}^{\tau\tau'}(q_{\rm eff})+R_{\Sigma''}^{\tau\tau'} W_{\Sigma''}^{\tau\tau'}(q_{\rm eff})+R_{\Sigma'}^{\tau\tau'} W_{\Sigma'}^{\tau\tau'}(q_{\rm eff})\Big] \\
+ \frac{q_\mathrm{eff}^2}{m_N^2}&\Big[R_{\Phi''}^{\tau\tau'}W_{\Phi''}^{\tau\tau'}(q_\mathrm{eff})+R_{\tilde{\Phi}'}^{\tau\tau'}W_{\tilde{\Phi}'}^{\tau\tau'}(q_\mathrm{eff})+R_{\Delta}^{\tau\tau'}W_{\Delta}^{\tau\tau'}(q_\mathrm{eff})\Big] \\
- \frac{2q_\mathrm{eff}}{m_N}&\Big[ R_{\Phi''M}^{\tau\tau'}W_{\Phi''M}^{\tau\tau'}(q_\mathrm{eff}) + R_{\Delta\Sigma'}^{\tau\tau'}W_{\Delta\Sigma'}^{\tau\tau'}(q_\mathrm{eff}) \Big]
\Bigg\},
\end{split}
\eeq
\begin{align}
&R_{M}^{\tau\tau'} =c_1^\tau c_1^{\tau'*}+c_{11}^\tau c_{11}^{\tau'*}, 
&&R_{\Sigma''}^{\tau\tau'} =\big(c_4^\tau-c_6^\tau\big) \big(c_4^{\tau'}-c_6^{\tau'}\big)^*+c_{10}^\tau c_{10}^{\tau'*},\nonumber \\
&R_{\Sigma'}^{\tau\tau'} =c_4^\tau c_4^{\tau'*}+c_9^\tau c_9^{\tau'*}, 
&&R_{\Phi''}^{\tau\tau'}=c_3^{\tau}c_3^{\tau' *}+(c_{12}^\tau - c_{15}^\tau)(c_{12}^{\tau' *}- c_{15}^{\tau' *}),\nonumber \\
&R_{\tilde{\Phi}'}^{\tau\tau'}=c_{12}^{\tau}c_{12}^{\tau' *}+c_{13}^\tau c_{13}^{\tau' *},
&&R_{\Delta}^{\tau\tau'}=c_{5}^{\tau}c_{5}^{\tau' *}+c_{8}^\tau c_{8}^{\tau' *},\nonumber \\
&R_{\Phi'' M}^{\tau\tau'}=\mathrm{Re}[c_3^\tau c_1^{\tau' *}-(c_{12}^\tau - c_{15}^\tau)c_{11}^{\tau' *}], 
&&R_{\Delta\Sigma'}^{\tau\tau'}=\mathrm{Re}[c_5^\tau c_4^{\tau' *}+c_8^\tau c_9^{\tau' *}],
\end{align}
where $\phi^{Z_\mathrm{eff}}_{1s}(\vec{0})$ is the $1s$ wave function of a muonic atom with effective charge $Z_{\rm eff}$, evaluated at the origin. This result factors the rate into a product of leptonic and nuclear response functions.  The leptonic responses $R_i^{\tau\tau'}$, bilinears in the NRET LECs,
determine the combinations of CLFV LECs that can in principle be isolated by measuring $\mu \rightarrow e$ rates in nuclear targets. The nuclear responses $W_i^{\tau\tau'}$, defined explicitly in \cite{ER2} in terms of sums over transition probabilities associated with multipole operators of rank $J$, can be thought of as experimental ``dials" that, by selecting targets with different ground-state properties, can be adjusted to isolate  the effects of specific $R_i^{\tau\tau'}$s.

The form of the rate in Eq. (\ref{eq:Gamma:mutoe}) can be derived from symmetry principles
alone, showing that the NRET amplitude obtained at order $\vec{v}_N$ is
general.  The NRET can be extended to include the additional operators
linear in $\vec{v}_\mu$, thereby accounting for contributions from the muon's lower component, in which case  \cite{ER2}
\begin{equation}
   \mathcal{L}_\mathrm{eff}^{\rm NRET} \rightarrow \sum_{N=n,p} \sum_{~i=1}^{16} c_i^N \mathcal{O}_i^N + \sum_{N=n,p}\sum_{~i\in 1 }^{10} b_i^N\mathcal{O}^{f,N}_i,
    \label{eq:L_NRET:vmu}
\end{equation}
but the net effect of the additional operators, for targets like $^{27}$Al, is to induce changes in
the nuclear response functions on the order of 5\%.

The embedding of the CLFV operators in a nucleus affects the CLFV physics that can be probed.
For free nucleons, the responses $W_{M}^{\tau\tau}$, $W_{\Sigma'}^{\tau\tau'}$, and $W_{\Sigma''}^{\tau\tau'}$ --- the charge, transverse spin, and longitudinal spin operators --- are $O(1)$, while contributions associated with the velocity-dependent responses $W_{\Phi''}^{\tau\tau'}$, $W_{\tilde{\Phi}'}^{\tau\tau'}$, and $W_{\Delta}^{\tau\tau'}$ are suppressed
by $q^2/m_N^2\sim m_\mu^2/m_N^2$.  This dependence is apparent in the momentum dependence of Eq. (\ref{eq:Gamma:mutoe}).

However, the nuclear embedding introduces additional effects due to operator coherence and to
selection rules imposed by parity and time-reversal symmetry on elastic interactions with the nucleus.
It is well known that the isoscalar charge response $W_{M}^{00}$ scales like $A^2$ (though with
some loss of strength due to the elastic form factor).  
However, the response $W_{\Phi''}^{\tau\tau'}$ also exhibits coherence
in nuclei like $^{27}$Al where one of two spin-orbit 
partner shells $j=\ell\pm 1/2$ is occupied \cite{Fitzpatrick:2012ix}. The response $\Phi''$, which is generically associated with tensor mediators, corresponds to the longitudinal projection of the nuclear spin-velocity current $\vec{v}_N\times\vec{\sigma}_N$.  In an example discussed in \cite{tower},
where the CLFV arises from leptoquark interactions, this coherence may have a significant impact on the overall rate, depending on the CLFV leptoquark couplings. The nucleus can also suppress certain interactions.  Examples include interactions associated
with a conserved vector current or an axial charge operator, with the latter vanishing due to
the combined constraints of parity and time-reversal invariance on elastic responses. Summarizing
the more extended discussion available in \cite{ER2}, such considerations yield, for isoscalar
coupling, the hierarchy
\beq
\label{eq:nuclear:response:scalings}
W_{M}^{00}\sim {O}(A^2)\gg \frac{q_{\rm eff}}{m_N}W_{M\Phi''}^{00}\gg \Big\{W_{\Sigma'}^{00},W_{\Sigma''}^{00},\frac{q_{\rm eff}^2}{m_N^2}W_{\Phi''}^{00} \Big\}\gg\Big\{\frac{q_{\rm eff}^2}{m_N^2}W_{\Delta}^{00},\frac{q_{\rm eff}^2}{m_N^2}W_{\tilde{\Phi}'}^{00}\Big\}.
\eeq

\begin{figure}[t]
    \centering
       \includegraphics[width=0.99\textwidth]{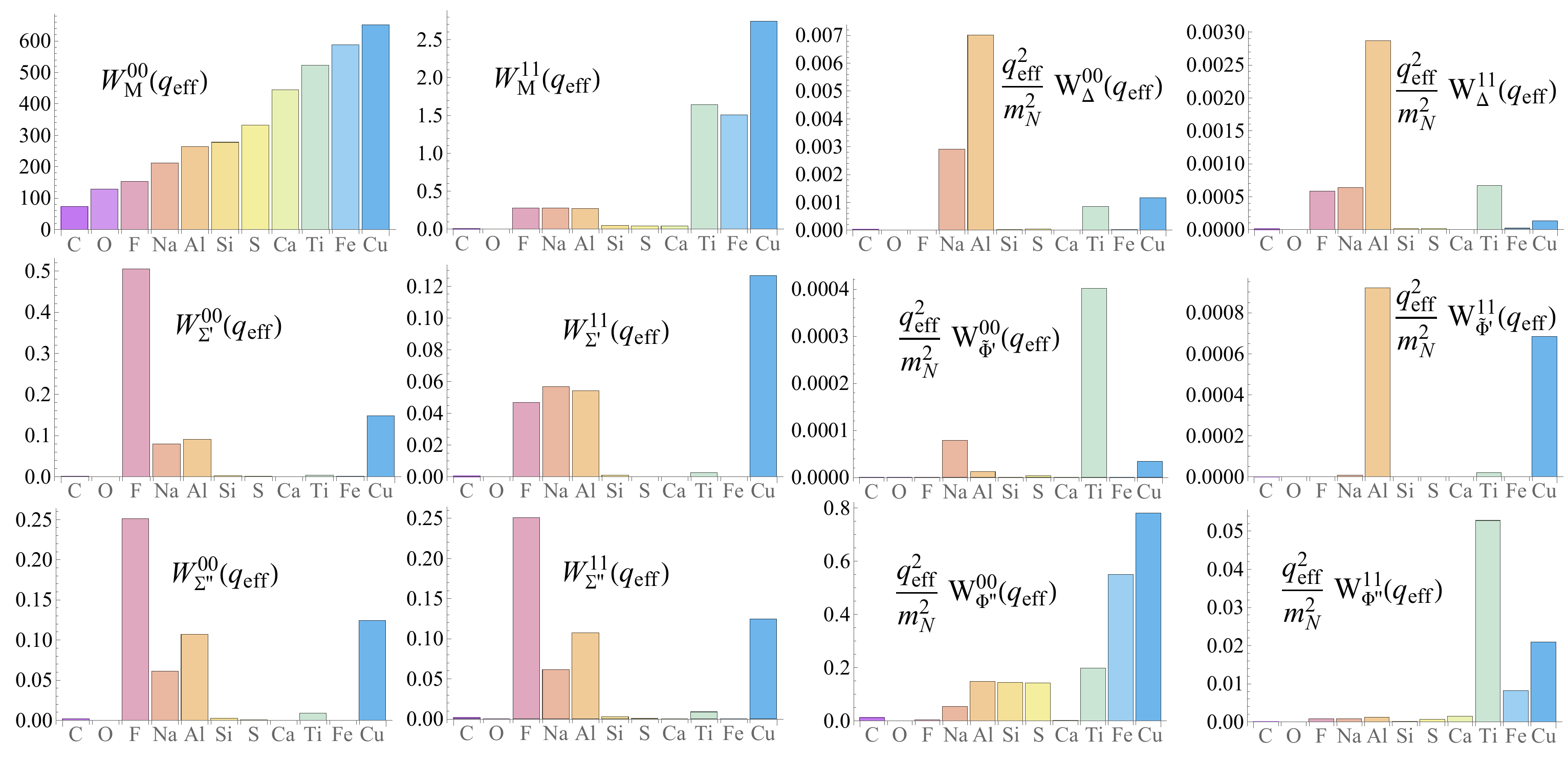}
    \caption{Illustration of the differences in target responses --- C, O, F, Na, Al, Si, S, Ca, Ti, Fe, Cu --- to allowed (left two panels) and velocity-dependent (right panels) operators.}
    \label{fig:Response}
\end{figure}

Figure \ref{fig:Response} summarizes the significant differences in target responses to possible CLFV operators. To illustrate how this sensitivity can be exploited, suppose CLFV were to be observed in a $J=0$, $T=0$
nucleus like $^{28}$Si, indicating a nonzero $W_{M}^{00}$ or $W_{\Phi^{\prime \prime}}^{00}$.  Both sources of coherence discussed above could contribute in $^{28}$Si.
If a second measurement done with the closed-shell nucleus $^{40}$Ca were to yield an even stronger signal, 
one would conclude that the CLFV source is the vector charge --- the operators associated with $c_1$ and $c_{11}$.
Conversely, if the second measurement yielded a weaker signal, the spin-vector response --- operators associated to
$c_3$, $c_{12}$, and $c_{15}$ --- would be indicated.

\begin{table*}[!]
\caption{Limits on the dimensionless weak-scale CLFV LECs $|\tilde{c}_i^\tau|$ and energy scale $\tilde{\Lambda}_i^\tau \equiv \Lambda_i^\tau/\mathrm{TeV}$,
from the indicated $\mu \rightarrow e$ branching ratios. 
Anticipated branching ratios are indicated by $^\dagger$. }
\label{tab:LEClimits}
\begin{tabular}{|c|l|l|l|H lH|}
\hline
 & & & & & &  \\[-7.5pt]
Target  &~~~~~~~~~~ Al&~~~~~~~~~~C&~~~~~~~~SiC&$^{32}$S~~~~~&~~~~~~~~~Ti&Cu\\[1.6pt]
\hline 
 & & & & & &  \\[-7.5pt]
 \diaghead{\theadfont DColumnnnHead II}%
 {Limits \\$|\tilde{c}^\tau_i|$ / $\tilde{\Lambda}^\tau_i$   }{Branch~\\ratio~\\} &  \thead{$10^{-17 \, \dagger}$} & \thead{$10^{-13 \, \dagger}$} & \thead{$2 \times 10^{-14 \, \dagger}$} & \thead{$7 \times 10^{-11}$} & \thead{$6.1 \times 10^{-13}$\cite{Wintz:1998rp}} & \thead{$1.6 \times 10^{-8}$\cite{Wintz:1998rp}} \\
 & & & & & &  \\[-7.5pt]
\hline
 & & & & & &  \\[-5.5pt]
$i=1,11~\tau=0$ &\,4.0$\cdot 10^{-10}$/1.2$\cdot 10^4$\,& \,5.1$\cdot 10^{-8}$/1.1$\cdot 10^3$\,&\,1.8$\cdot 10^{-8}$/1.9$\cdot 10^3$\,&\,1.0$\cdot 10^{-6}$/240\,&\,7.4$\cdot 10^{-8}$/910\,&\,1.2$\cdot 10^{-5}$/\,71\,\\
$i=1,11~\tau=1$ &\,1.2$\cdot 10^{-8~}$/2.2$\cdot 10^3$\,& \,6.3$\cdot 10^{-6}$/~~~98\,&\,1.4$\cdot 10^{-6}$/$~~210$\,& - &\,1.3$\cdot 10^{-6}$/$210$\,&\,1.9$\cdot10^{-4}$/\,18\,\\
$i=3,15 ~\tau=0$ &\,1.6$\cdot 10^{-8~}$/1.9$\cdot 10^3$\,& \,4.0$\cdot 10^{-6}$/$~~120$\,&\,7.3$\cdot 10^{-7}$/$~~290$\,&\,4.5$\cdot 10^{-5}$/\,37\,\,&\,3.8$\cdot 10^{-6}$/130\,&\,3.5$\cdot10^{-4}$/\,13\,\\
$i=3,15 ~\tau=1$ &\,1.9$\cdot 10^{-7~}$/$~~570$\,& \,1.3$\cdot10^{-4}$/~~~21\,&\,4.0$\cdot 10^{-5}$/~~~39 \, & - &\,7.3$\cdot 10^{-6}$/~91\,&\,2.1$\cdot 10^{-3}$/5.3\,\\
$i=4~~~~~ \tau=0$ &\,1.4$\cdot 10^{-8~}$/2.1$\cdot 10^3$\,& \,9.4$\cdot 10^{-6}$/~~~80 \,&\,4.1$\cdot 10^{-6}$/$~~120$\,& - &\,1.5$\cdot 10^{-5}$/~63\,&\,5.9$\cdot10^{-4}$/\,10\,\\
$i=4 ~~~~~ \tau=1$ &\,1.7$\cdot 10^{-8~}$/1.9$\cdot 10^3$\,& \,1.1$\cdot 10^{-5}$/~~~76 \,&\,4.9$\cdot 10^{-6}$/$~~110$\,& - &\,1.7$\cdot 10^{-5}$/~59\,&\,6.1$\cdot10^{-4}$/9.9\,\\
$i =5,8 ~~ \tau=0$  &\,7.8$\cdot 10^{-8~}$/$~~880$\,& \,9.6$\cdot 10^{-5}$/~~~25\,&\,7.1$\cdot 10^{-5}$/~~~29 \,& - &\,5.8$\cdot 10^{-5}$/~32\,&\,9.0$\cdot 10^{-3}$/2.6\,\\
$i=5,8 ~~ \tau=1$  &\,1.2$\cdot 10^{-7~}$/$~~720$\,& \,1.6$\cdot10^{-4}$/~~~20\,&\,7.3$\cdot 10^{-5}$/~~~29 \,& - &\,6.5$\cdot 10^{-5}$/~30\,&\,2.7$\cdot 10^{-2}$/1.5\,\\
$i=6,10~ \tau=0$  &\,2.0$\cdot 10^{-8~}$/1.8$\cdot 10^3$\,& \,1.1$\cdot 10^{-5}$/~~~75\,&\,5.5$\cdot 10^{-6}$/$~~110$\,& - &\,1.8$\cdot 10^{-5}$/~59\,&\,8.7$\cdot10^{-4}$/8.3\,\\
$i=6,10~\tau=1$  &\,2.2$\cdot 10^{-8~}$/1.7$\cdot 10^3$\,& \,1.2$\cdot 10^{-5}$/~~~71\,&\,6.1$\cdot 10^{-6}$/~~~99 \,& - &\,2.0$\cdot 10^{-5}$/~55\,&\,8.7$\cdot10^{-4}$/8.3\,\\
$i=9~~~~~\tau=0$   &\,2.1$\cdot 10^{-8~}$/1.7$\cdot 10^3$\,& \,1.9$\cdot 10^{-5}$/~~~57\,&\,6.2$\cdot 10^{-6}$/~~~99 \,& - &\,2.8$\cdot 10^{-5}$/~47\,&\,8.0$\cdot10^{-4}$/8.7\,\\
$i=9~~~~~\tau=1$  &\,2.8$\cdot 10^{-8~}$/1.5$\cdot 10^3$\,& \,2.2$\cdot 10^{-5}$/~~~52\,&\,8.1$\cdot 10^{-6}$/~~~87 \,& - &\,3.4$\cdot 10^{-5}$/~42\,&\,8.7$\cdot10^{-4}$/8.4\,\\
$i=12~~~~\tau=0$ &\,1.6$\cdot 10^{-8~}$/1.9$\cdot 10^3$\,& \,4.0$\cdot 10^{-6}$/$~~120$\,&\,7.3$\cdot 10^{-7}$/$~~290$\,&\,4.5$\cdot 10^{-5}$/\,37\,\,&\,3.8$\cdot 10^{-6}$/130\,&\,3.5$\cdot10^{-4}$/\,13\,\\
$i=12~~~~\tau=1$  &\,1.4$\cdot 10^{-7~}$/$~~660$\,& \,1.3$\cdot10^{-4}$/~~~21 \,&\,4.0$\cdot 10^{-5}$/~~~39 \,& - &\,7.3$\cdot 10^{-6}$/~91\,&\,2.1$\cdot 10^{-3}$/5.4\,\\
$i=13~~~~\tau=0$ &\,1.8$\cdot 10^{-6~}$/$~~180$\,&~~~~~~~~~~~ - &~~~~~~~~~~~ - & - &\,8.4$\cdot 10^{-5}$/~27\,&\,5.3$\cdot 10^{-2}$/1.1\,\\
$i=13~~~~\tau=1$ &\,2.1$\cdot 10^{-7~}$/$~~540$\,&~~~~~~~~~~~ - &~~~~~~~~~~~ - & - &\,3.7$\cdot10^{-4}$/~13\,&\,1.2$\cdot 10^{-2}$/2.3\,\\[2.6pt]
\hline
\end{tabular}
\end{table*} 

\section{Mu2e and COMET Sensitivities}
The resulting limits on CLFV are given in Table \ref{tab:LEClimits}, expressed 
in terms of the dimensionless weak-scale couplings $|\tilde{c}_i^\tau|$, where $c_i^\tau = \sqrt{2} G_F \tilde{c}_i^\tau=1/(\Lambda_i^\tau)^2$.
We have separately explored each of the 12 LECs that contribute to elastic $\mu \rightarrow e$ conversion, to assess the comparative
sensitivity of anticipated experiments.  The table shows the versatility of the $^{27}$Al target chosen by Mu2e and
COMET: the ground-state angular momentum of $\frac{5}{2}^+$ allows each of the 12 operators to contribute. As a $T=\frac{1}{2}$ nucleus, $^{27}$Al is sensitive to both
isoscalar and isovector operators (though, as an odd-proton nucleus, it is nearly blind to spin-dependent
interactions coupling only to neutrons).   See \cite{ER1,ER2} for additional discussion.

\section{Inelastic $\mu \rightarrow e$ Conversion}
Given a discovery of $\mu \rightarrow e$ conversion, it would appear that the signal could in principle be attributed to any one of the interactions $\CO_i$,
as nothing is known about the BSM source of the CLFV.  However, this is not the case for an experiment employing $^{27}$Al.  In addition to the $5/2^+$ ground
state, this nucleus has three excited states of interest, the $1/2^+$, $3/2^+$, and $7/2^+$ at 0.844, 1.015, and 2.212 MeV, respectively. The electron background
from radiative muon capture does not contribute below 3.62 MeV.  The SM decay-in-orbit (DIO)
background ($\mu \rightarrow e + \bar{\nu}_e + \nu_\mu$) is sufficiently low and its shape sufficiently well understood that inelastic $\mu \rightarrow e$ conversion to these three states can be resolved under a shape analysis.  The background analysis, based on Mu2e simulations \cite{universe9010054}, is described in more detail in \cite{Haxton:2024ecp,PhysRevC.111.025501}.

\begin{figure}[t]
    \centering
    \includegraphics[width=0.8\textwidth]{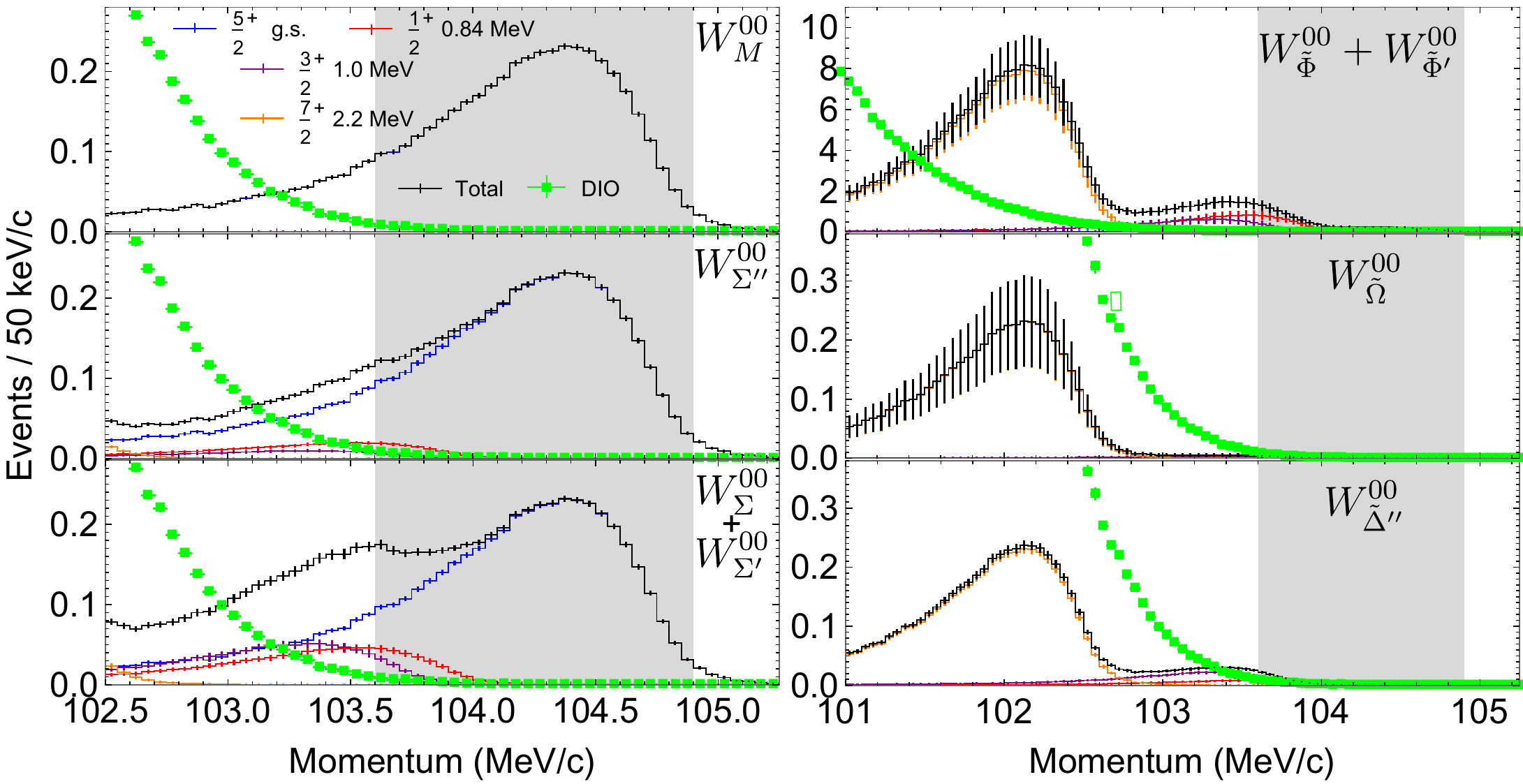}
    \caption{Expected electron counts (50 keV/c bins) in  Mu2e Run I for CLFV mediated by isoscalar charge ($M$), longitudinal spin ($\Sigma^{\prime \prime}$), transverse
    spin ($\Sigma$ + $\Sigma^\prime$), transverse spin-velocity ($\tilde{\Phi}$ and $\tilde{\Phi}^\prime$), axial charge ($\tilde{\Omega}$), and longitudinal convection ($\tilde{\Delta}^{\prime \prime}$)
    interactions.  The total signal (black) is obtained by combining ground state (blue) and exited state [0.84 MeV (red), 1.0 MeV (purple), and 2.2 MeV (orange)] contributions.  The green squares are
    the estimated DIO background.  The gray shading indicates the energy interval where the Mu2e sensitivity is optimized.  The error bars are estimates of the combined uncertainty from
    statistics and from the nuclear physics of the responses.
    \label{fig:inelastic}}
\end{figure}

While the form of the single-nucleon NRET is general, the NRET operators and response functions contributing to the elastic nuclear rate of Eq. \eqref{eq:Gamma:mutoe} are constrained by time-reversal. This constraint is relaxed for inelastic transitions, with NRET interactions like $\CO_2$, $\CO_7$, $\CO_{14}$, and $\CO_{16}$ now contributing along with new
nuclear response functions. This inelastic generalization is described in \cite{Haxton:2024ecp,PhysRevC.111.025501} and will not be repeated here.  The implications of these papers for Mu2e are summarized in Fig. \ref{fig:inelastic}, where the expected electron signal for various CLFV scenarios is shown (assuming a ground-state branching ratio $B(\mu\rightarrow e) = 10^{-15}$). These Run-I simulations use only $\sim$6\% of the expected total data set.  It is clear that Mu3e and COMET not only have the potential to discover CLFV, but also to provide
a great deal of information about the operator mechanism.  Even the absence of an excited-state contribution provides significant new information.\\

\noindent
{\it Acknowledgments:} The authors are supported by the U.S. Department of Energy under grants DE-SC0004658 and DE-AC02-05CH11231 and (ER)
by Los Alamos National Laboratory, operated by Triad National Security, LLC, for the National Nuclear Security Administration of U.S. Department of Energy (Contract No. 89233218CNA000001).
Additional support was provided by the National Science Foundation under cooperative agreement 2020275.  


\end{document}